\documentclass[seceq]{ptptex}

\usepackage{graphicx}
\usepackage{hyperref}

\newcommand{\beq}{\begin{equation}}
\newcommand{\eeq}{\end{equation}}
\newcommand{\beqa}{\begin{eqnarray}}
\newcommand{\eeqa}{\end{eqnarray}}

\def\vereq#1#2{\lower3.5pt\vbox{\baselineskip1.5pt \lineskip1.5pt
\ialign{$#1\hfill##\hfil$\crcr#2\crcr\sim\crcr}}}
\newcommand{\lsim}{\mathrel{\mathpalette\vereq<}}
\newcommand{\gsim}{\mathrel{\mathpalette\vereq>}}

\newcommand{\TeV}{{\rm TeV}}
\newcommand{\GeV}{{\rm GeV}}

\newcommand{\lqcd}{\Lambda_{\rm QCD}}

\newcommand{\Bbar}{\,\overline{\!B}{}}
\newcommand{\Dbar}{\,\overline{\!D}{}}
\newcommand{\Kbar}{\,\overline{\!K}{}}
\def\B0bar{\Bbar{}^0}
\def\D0bar{\Dbar{}^0}
\def\K0bar{\Kbar{}^0}

\def\rhobar{\bar\rho}
\def\etabar{\bar\eta}

\newcommand{\ov}{\overline}

\font\smallsl=cmsl8
\def\babar{\mbox{\sl B\kern-0.1em\hbox{\smallsl A}\kern-0.03em
  B\kern-0.1em\hbox{\smallsl A\kern-0.05em R}}}

\title{
Future prospects of $B$ physics%
}

\author{
Yuval \textsc{Grossman}$^{\,1}$, Zoltan \textsc{Ligeti}$^{\,2}$, and Yosef
\textsc{Nir}$^{\,3}$}

\inst{
$^1$Newman Laboratory of Elementary Particle Physics\\
        Cornell University, Ithaca, NY 14853, USA\\[2pt]
$^2$Ernest Orlando Lawrence Berkeley National Laboratory\\
        University of California, Berkeley, CA 94720, USA\\[2pt]
$^3$Department of Particle Physics, Weizmann Institute of Science\\
        Rehovot 76100, Israel}

\abst{In recent
years, the CKM picture of flavor and $CP$ violation has been
confirmed, mainly due to $B$ decay data. Yet, it is likely that there
are small corrections to this picture. We expect to find new physics
not much above the weak scale. This new physics could modify flavor
changing processes compared to their SM expectations. Much larger $B$
decay data sets, which are expected from LHCb and super-$B$-factories,
will be used to search for these deviations with much improved
sensitivity. The combination of low and high energy data will be
particularly useful to probe the structure of new physics.}

\begin{document}

\maketitle

\section{Introduction}

The aim of high energy physics is to understand the fundamental
interactions among the elementary particles in Nature. The
mathematical tool that is used to describe these interactions is
quantum field theory, and high energy physics aims to determine the
Lagrangian of Nature. The standard model (SM) has so far proven to be
a good description of Nature up to energies of the order of the
electroweak scale, $m_W\sim 100\,\GeV$.  In particular, the
electroweak gauge sector of the SM has been verified to an accuracy
better than $1\%$, mainly by the LEP, SLC, and Tevatron
experiments~\cite{Amsler:2008zzb}.

In recent years, our understanding of the flavor sector has improved
dramatically, due to the $e^+ e^-$ $B$ factories, \babar, Belle, and
CLEO, and the Tevatron experiments.  With over $10^9$ decays of $B$
hadrons analyzed, the SM picture of the flavor sector has been tested
with impressive accuracy \cite{Amsler:2008zzb}.

One sector where deviations from the SM predictions have been found is the
lepton sector. The experimental pieces of evidence for neutrino flavor
transitions contradict the prediction of massless neutrinos. The SM
can be modified in a simple way to accommodate this result. If we
consider the SM to be a low energy effective theory,
nonrenormalizable terms should be included in the SM action. In
particular, the only dimension-five operators that can be added
generate neutrino masses.  The observed coefficients of these
operators suggest a suppression scale that is very high, well beyond
direct probe.

The least tested sector of the SM is the Higgs sector.  The Higgs
boson is the only SM particle that has not yet been discovered. There are
numerous alternatives to the SM Higgs sector. These range from mild
modifications, such as having additional Higgs doublets, to rather radical
alternatives where the mechanism that breaks the electroweak symmetry is not the
vacuum expectation value of an elementary scalar.
One of the most disturbing aspects of the SM is related to the Higgs sector.
Once the SM is viewed as an effective theory, the fact that the Higgs mass is
not very high becomes puzzling: radiative corrections involving heavy particles
would drive the Higgs mass close to the cutoff scale.
This well known fine tuning problem is often interpreted as an indication for
new physics at the weak scale. Indeed, this is the main motivation to look for
new effects at the weak scale with the LHC.

Flavor physics, in particular $B$ physics, has provided strong upper 
bounds on contributions from new physics models.  This situation leads to
the ``new physics flavor puzzle", which is the mismatch between the 
relatively low scale required to solve the fine tuning problem,
and the high scale that is seemingly required to suppress the non-SM
contributions to flavor changing processes. Let us expand a little
on this point. The flavor sector of the SM is impressively
successful.  This success is linked to the fact that the SM flavor structure is
special. First, the CKM matrix is unitary and contains small mixing angles.
Second, flavor-changing neutral currents (FCNCs) are highly suppressed.  These
features are crucial to explain the observed pattern of weak decays.  Any
extension of the SM must preserve these successful features.
Consider a model where the only suppression of new flavor changing interactions
comes from the large masses (of scale $\Lambda\gg m_W$) of the new particles
that mediate them. Flavor physics, in particular measurements of meson mixing
and $CP$ violation, put severe lower bounds of order $\Lambda\gsim 10^4\,\TeV$.
There is therefore a tension.  The hierarchy problem can be solved with new
physics at a scale $\Lambda \sim 1$\,TeV.  Flavor bounds, on the other hand,
require $\Lambda \gsim 10^4$\,TeV.  This tension implies that any TeV-scale new
physics cannot have a generic flavor structure. The new physics flavor
puzzle is thus the question of why, and in what way, the flavor
structure of the new physics is non-generic.

Flavor physics has been mainly an input to model building, not an
output. The flavor predictions of most new physics models are not a
consequence of their generic features but rather of the special
structure that is imposed specifically to satisfy the existing severe
flavor bounds. Therefore, flavor physics is a powerful indirect probe
of new physics. We hope that new physics not far above the weak scale
will be discovered at the LHC. A major issue will then be to
understand its flavor structure. While it is not easy to directly
probe this flavor structure at high energy, a lot can be learned
from low energy flavor physics.

The precision with which we can probe the high scale physics in flavor physics
experiments is limited by theoretical uncertainties (once experimental
precision becomes good enough). Thus, the important questions are the
following:
\begin{enumerate}
\item
What are the expected deviations from the SM predictions induced by
new physics at the TeV scale?
\item
What are the theoretical uncertainties?
\item
What can we expect in terms of experimental precision?
\item
What will the measurements teach us if deviations from the SM are [not] seen?
\end{enumerate}
In the following we  discuss these questions in detail. The main lines of
our answers read as follows:
\begin{enumerate}
\item
The expected deviations from the SM predictions induced by new physics at the
TeV scale with generic flavor structure are already ruled out by many
orders of magnitudes. We can thus expect any size of deviation below the
current bounds. In a large class of scenarios we expect deviations at
the $10^{-2}$ level.
\item
The theoretical limitations are highly process dependent. Some measurements are
already limited by theoretical uncertainties (mostly due to hadronic, strong
interaction, effects), while in various other cases the theory has very small
uncertainties, and is more precise than the expected sensitivity of
future experiments.
\item
Experimentally, the useful data sets can increase by a factor of order one
hundred at LHC-b and a super-$B$ factory. Such improvements will therefore
probe into the region of fairly generic new physics predictions.
\item
The new low energy flavor data will be complementary with the high-$p_T$ 
part of the LHC program. The synergy of both data sets can teach us a lot 
about the new physics at the TeV scale.
\end{enumerate}

In the next section we briefly review the current status of (quark) flavor
physics.  Sections 3--6 discuss questions 1--4, respectively.  Section 7
contains our conclusions.

\section{Current status}

In the standard model, the distinction between quarks of different
generations comes from their Yukawa couplings to the Higgs field. In
the mass basis, this flavor physics is manifest in quark masses, in
$CP$ violation, and in all flavor changing phenomena described by the
Cabibbo-Kobayashi-Maskawa (CKM) quark mixing matrix~\cite{C,KM}.  In
the SM all flavor-changing phenomena are described by only a handful
of parameters, and therefore intricate correlations are predicted
between dozens of different decays of $s$, $c$, $b$, and $t$ quarks,
and in particular between $CP$ violating observables.  Possible
deviations from the CKM paradigm may modify (i) correlations between
various measurements (e.g., inconsistent constraints from $B$ and $K$
decays, or from $CP$ asymmetries in different decay modes, for
example, $B\to \psi K$ and $B \to \phi K$); (ii) predictions for FCNC
transitions (e.g., enhanced $B_{(s)}\to \ell^+ \ell^-$); (iii)
enhanced $CP$ violation, (e.g., in $B\to K^*\gamma$ or in $B_s\to \psi
\phi$).

Over the past decade, much progress has been made in precision measurements
of the flavor parameters and in testing the SM flavor sector by many
overconstraining measurements.  To visualize the constraints from many
measurements, it is convenient to use the Wolfenstein
parameterization~\cite{Wolfenstein:1983yz} of the CKM matrix,
\beq\label{ckmdef}
V = \left( \begin{array}{ccc}
  V_{ud} & V_{us} & V_{ub} \\
  V_{cd} & V_{cs} & V_{cb} \\
  V_{td} & V_{ts} & V_{tb} \end{array} \right)
= \left( \begin{array}{ccc}
  1-\frac{1}{2}\lambda^2 & \lambda & A\lambda^3(\rhobar-i\etabar) \cr
  -\lambda & \!\!1-\frac{1}{2}\lambda^2\!\! & A\lambda^2 \cr
  A\lambda^3(1-\rhobar-i\etabar) & -A\lambda^2 & 1 \end{array} \right)
  + \ldots ,
\eeq
This parameterization exhibits the hierarchical structure of the CKM matrix by
expanding in a small parameter, $\lambda \simeq 0.23$; however, recent CKM fits
use definitions of the $\lambda,\, A,\, \rhobar$ and $\etabar$ parameters that
obey unitarity exactly~\cite{Charles:2004jd}. The unitarity of $V$ implies
\beq
\sum_i V_{ij} V_{ik}^* = \sum_i V_{ji} V_{ki}^* = \delta_{jk}.
\eeq
Each of the six vanishing combinations can be represented by a triangle in
the complex plane.  The most commonly used such triangle, often
called ``the unitarity triangle,'' arises from rescaling the
\beq \label{The-UT}
V_{ud}\, V_{ub}^* + V_{cd}\, V_{cb}^* + V_{td}\, V_{tb}^* = 0
\eeq
relation by $V_{cd}\,V_{cb}^*$ and choosing two vertices of the
resulting triangle to be $(0,0)$ and $(1,0)$.  The definition
\beq
\rhobar + i \etabar = -\frac{V_{ud}V_{ub}^*}{ V_{cd}V_{cb}^*},
\eeq
ensures that the apex of the unitarity triangle is $(\rhobar,\etabar)$.

\begin{figure}[t!]
\centerline{\includegraphics*[height=4.5cm]{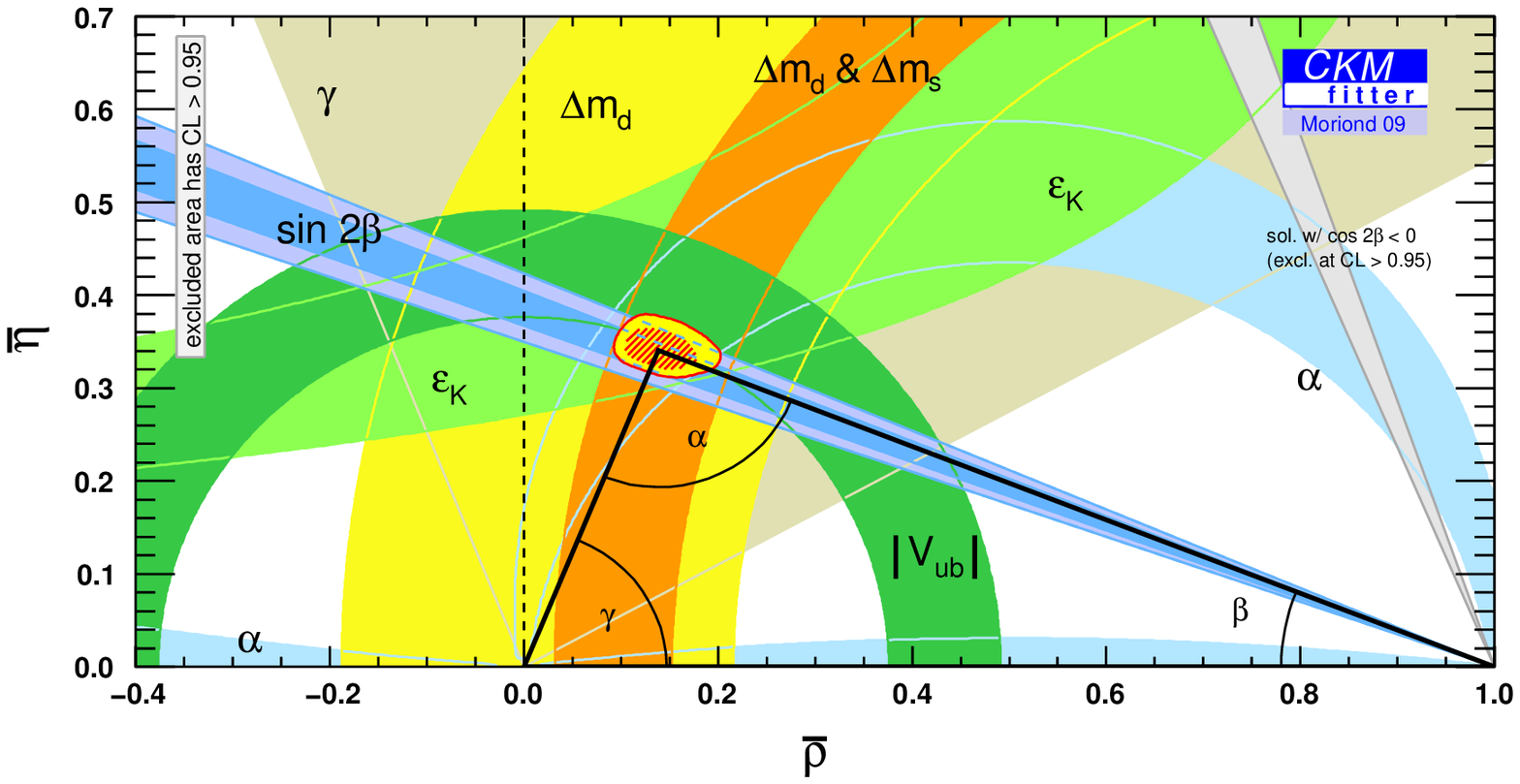} \hfill
\raisebox{4pt}{\includegraphics*[height=4.5cm]{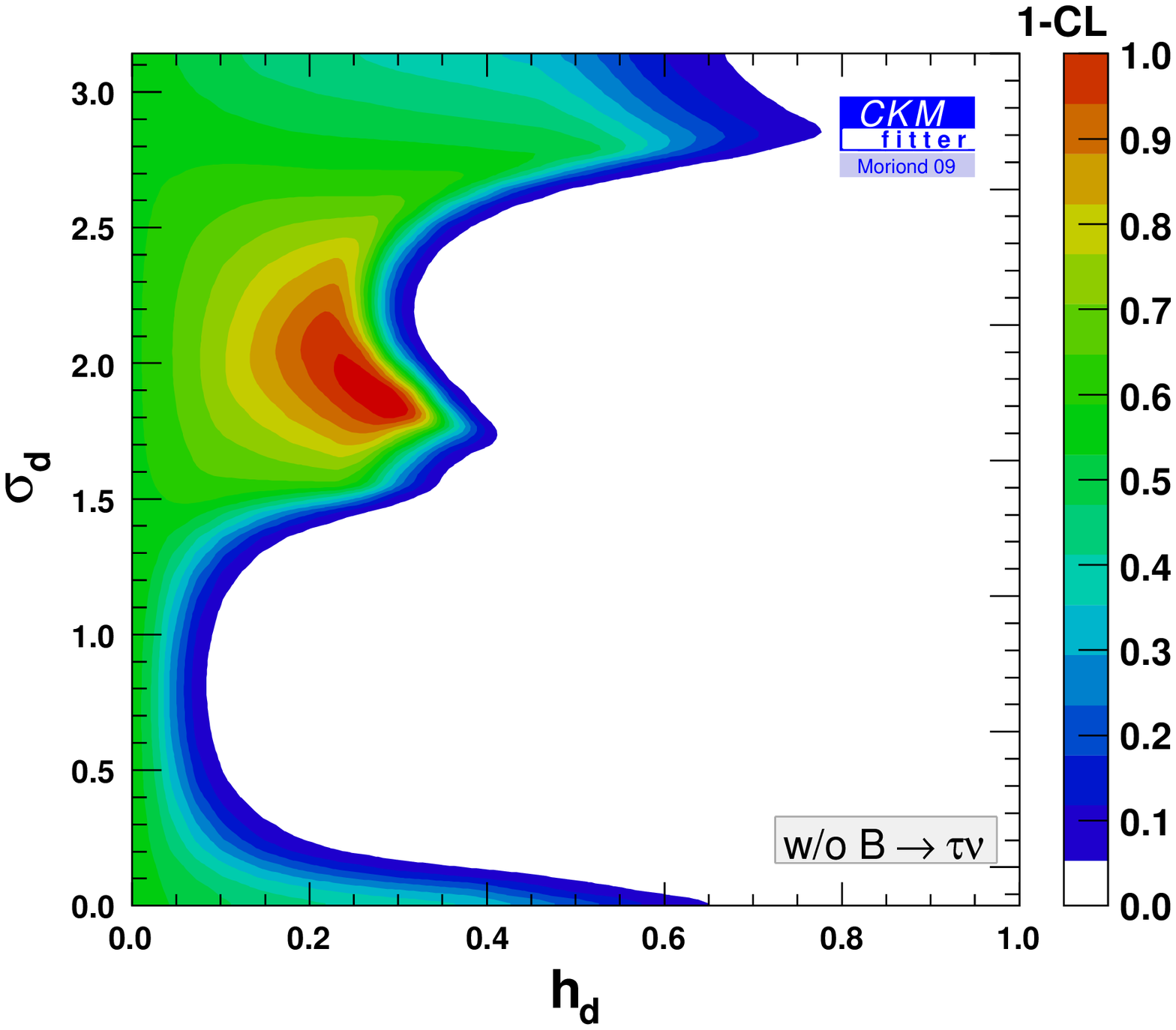}}}
\vspace*{-.25cm}
{\footnotesize\hspace*{2cm} (a)\hspace*{7.5cm}(b)}
\caption{Left: Constraints on the apex of the unitarity triangle in the
$\rhobar-\etabar$ plane.  The shaded regions show the 95\%\,CL
constraints.  Right: the allowed $h_d-\sigma_d$ region in the presence of new
physics in $B^0$--$\B0bar$ mixing, as parameterized by
Eq.~(\ref{hsigma}).~{\cite{ckmfitter,Charles:2004jd}}}
\label{fig:hdsd}
\end{figure}

The asymmetric-energy $B$ factory experiments, \babar\ and Belle, have
measured many $CP$ violating observables (around 20 with more than
$3\sigma$ significance), of which the most precise is the $CP$
asymmetry in $B\to \psi K_S$ and related modes,
\beq
S_{\psi K} = \sin2\beta = +0.671 \pm 0.024\,,
\eeq
with only a 4\% experimental uncertainly.\footnote{In the literature
there are two common ways to refer to the angles of the unitarity
triangle. We use the $\alpha$, $\beta$ and $\gamma$ notation. The
other notation is, respectively, $\phi_2$, $\phi_1$, and $\phi_3$.} As
shown in Fig.~\ref{fig:hdsd}(a), the result of this and other
measurements is that the $CP$ violating parameter $\etabar$ (or,
equivalently, the Jarlskog invariant, $J$) has been determined with a
5\% uncertainty.  This figure indicates that the existing measurements
are consistent with the CKM picture of quark mixing, and, in particular,
with the KM phase~\cite{KM} being responsible for the observed $CP$ violation.

The implications of the level of agreement between these various
measurements is often overstated, however.  Fig.~\ref{fig:hdsd}(a) does not
directly address how well the existing measurements constrain
additional non-SM contributions to flavor changing processes.  This
cannot be easily discussed in a model independent fashion, since the most 
general low energy effective Hamiltonian contains about a hundred
dimension-six operators, which parameterize the leading corrections to
the SM.  In a large class of models, however, the new physics modifies 
the mixing amplitude of neutral mesons, while
leaving tree level decays unaffected. This effect can be parameterized
with just two real parameters for each mixing amplitude.
For $B^0 - \B0bar$ mixing, we write
\beq\label{hsigma}
M_{12} = M_{12}^{\rm SM}\, \big(1 + h_d\, e^{2i\sigma_d}\big)\,.
\eeq
Fig.~\ref{fig:hdsd}(b) shows the constraints on $h_d$ and $\sigma_d$,
indicating that order $10-20\%$ corrections to $|M_{12}|$ are still
allowed, for (almost) any value of the phase of the new physics contribution.
If this phase is aligned with the SM, $2\sigma_d = 0$ mod $\pi$, then
the new physics contribution to $|M_{12}|$ may still be comparable to the SM
one.  Similar conclusions apply to new physics contributions to many other
FCNC transition amplitudes, though those analyses tend to be more
complicated and (strong interaction) model dependent.

The fact that such large deviations from the SM are not yet excluded
gives a very strong motivation to continue low energy flavor
measurements in order to observe deviations from the SM predictions or
establish a stronger hierarchy between the SM and new physics contributions.

\section{Typical effects of TeV-scale new physics}

There are at least two good reasons to think that there is new physics
at (or below) the TeV scale. First, the fine-tuning problem of the
Higgs mass implies that either there are some symmetry partners of (at
least) the particles that couple to the Higgs with order one couplings
(the top quark, the $W$- and $Z$-bosons, and the Higgs itself), or
gravity changes its nature at this scale. Second, if the dark matter
particles are weakly interacting massive particles (WIMPs), then their
annihilation cross section is of order 1/TeV$^2$.  The first argument
for TeV-scale physics further implies that the new physics has 
nontrivial flavor structure, because some of its effects must be related
to the top Yukawa coupling. In this section we describe the flavor
effects expected from TeV-scale new physics.

The effects of new physics can be parameterized by
nonrenormalizable operators, made of the standard model fields,
obeying the standard model gauge symmetries, and suppressed
by inverse powers of the scale of the new physics.
In particular, there is a large number of dimension-six
four-fermion operators that contribute to FCNCs. Let us take
as an example the following four-quark $\Delta B=2$ operator:
\beq\label{zbdope}
\frac{z_1^{bd}}{\Lambda_{\rm NP}^2}\, (\overline{d_L}\gamma_\mu b_L)^2,
\eeq
where $z_1^{bd}$ is a dimensionless coefficient and $\Lambda_{\rm NP}$
is the scale of new physics. This operator contributes to
$B^0-\overline{B^0}$ mixing. Its absolute value is then constrained by
$\Delta m_B$ and its imaginary part (in the basis where the leading
decay amplitudes are real) by $S_{\psi K_S}$. For a detailed analysis,
the reader is referred to Ref. \cite{Bona:2007vi}.

Let us now focus on
\beq
\Lambda_{\rm NP}\lsim1\;\TeV.
\eeq
If the contribution of the new physics comes at tree level,
with complex couplings of order one, then
the contribution of (\ref{zbdope}) will be 5--6 orders of
magnitude above the experimental bounds. If the contribution
comes at the loop level, say ${\cal O}(\alpha_2^2)$ as in
the standard model, but there are no additional suppression
factors, then the contribution is still 2--3 orders of magnitude
above the experimental bounds. The contribution from operators with a
different Lorentz structure, in particular
\beq\label{zbdfour}
\frac{z_4^{bd}}{\Lambda_{\rm NP}^2}\, (\overline{d_R}\, b_L)
  (\overline{d_L}\,b_R),
\eeq
is even two orders of magnitude larger (for the same loop 
suppression factors). These facts lead to two important conclusions:
\begin{itemize}
\item New physics at the TeV scale can easily saturate the upper
bounds on FCNC. Conversely, even a mild improvement in the
experimental sensitivity may lead to signals of
new physics in $B$ decays.
\item New physics at the TeV scale must have a highly
non-generic flavor structure. This special structure might
involve alignment, particularly with the down sector,
or some level of flavor degeneracy, or a combination of the two.
\end{itemize}

For $B$ decays, and with ${\cal O}(10^{-3})$ flavor-blind suppression,
the flavor factors must suppress, for example, $z_1^{bd}$ and
$z_4^{bd}$ by order $10^{-3}$ and $10^{-5}$, respectively. Suppose
that the suppression of $z_1^{bd}$ comes from both a loop factor of
order $10^{-3}$ and flavor alignment similar in size to that of the
standard model:
\beq
z_1^{bd}\sim 10^{-3} (V_{tb}V_{td}^*)^2\sim10^{-7}.
\eeq
This is a few percent effect in $B^0-\overline{B^0}$ mixing, which is
too small to be observed in $\Delta m_B$. If, however, the new phase
carried by $z_1^{bd}$ is of order one, then the effect on $S_{\psi K_S}$
is also of order a few percent, which may be observed in the future.
Note that, in general, we do not expect flavor degeneracy in $z_1^{bd}$
because the global $SU(3)_Q$ symmetry is strongly broken even within
the standard model. The analogous naive estimate for $z_4^{bd}$, based
on supersymmetric models of alignment \cite{Nir:1993mx}, gives
\beq
z_4^{bd}\sim 10^{-3}\, (m_d/m_b)\sim10^{-6},
\eeq
which can saturate the bounds.

Finally, it is possible that $z_1^{bd}$ is not just of the order of
$(V_{tb}V_{td}^*)^2$, but actually proportional to it. This happens in
a class of models called minimal flavor violation (MFV).  An example
of such a model is low-energy gauge-mediated supersymmetry breaking.
Then the phase of $z_1^{bd}$ is the same as the one carried by the
standard model, and there will be no deviation from the standard model
prediction for $S_{\psi K_S}$. Furthermore, in this class of models,
flavor changing operators involving right-handed quarks, such as
(\ref{zbdfour}), are highly suppressed.

Similar considerations apply to other operators that contribute to
$B^0-\overline{B^0}$ mixing, as well as operators that contribute to
FCNC decay amplitudes, for example $b\to s\ell^+\ell^-$.  With new
physics at the TeV scale, the new contributions can be as large as the
present upper bounds. With loop and flavor suppression factors, there
could be effects of order a few percent, which can lead to observable $CP$
violating effects. Finally, if the new physics contributions are loop
suppressed and carry the CKM angles and phases, then the effects on
the theoretically cleanest $CP$ asymmetries vanish, while the
contributions to $CP$ conserving observables tend to be smaller
than the theoretical uncertainties.

\section{Theoretical limitations}

In order to have a convincing signal of new physics, we need its effect to
be larger than both the experimental and the theoretical uncertainties. The 
most interesting observables are thus those with small theoretical uncertainties
and good experimental precision. Since we search for deviations from the SM,
the relevant theoretical uncertainties are those of the SM predictions.

There are in general two kinds of theoretical uncertainties, usually
labeled perturbative and nonperturbative.  Perturbative uncertainties
come from the truncation of expansions in small (or not-so-small)
coupling constants, such as $\alpha_s(m_b) \simeq 0.2$.  There are
always higher order terms that have not been computed. In principle,
such errors can be reduced by performing the higher order
calculations; however, the calculations become very demanding. So far,
when the precision of a perturbative QCD calculation was the limiting
uncertainty, the necessary calculations were possible to carry out. A
prime example is the next-to-next-to-leading order calculation
(four-loop running, three-loop matching and matrix elements) of $B \to
X_s
\gamma$\cite{Misiak:2006zs}.

The second kind of uncertainty is due to nonperturbative effects.  They arise
because QCD becomes strongly  interacting at low energies, and one can no longer
expand in the coupling constant. In general, there is no systematic method to
deal with nonperturbative effects. There are cases, however, when we can get at
the fundamental physics even in the presence of such effects.  One way to
proceed is to find observables where all (or most) of the hadronic parameters
cancel or can be extracted from data.  Many interesting processes involve
hadronic parameters which can neither be measured nor calculated in perturbation
theory.  In some cases other tools can be used, such as lattice QCD or
exploiting symmetries of the strong interaction which arise in certain limits,
such as chiral symmetry or the heavy quark expansion.  Then the limiting
uncertainty is often due to the increasing number of hadronic matrix elements as
one goes to higher order. These methods often use experimental data from related
processes to constrain the uncertainties. Thus, experimental progress will not
only reduce the measurement errors, but can also reduce the theoretical
uncertainties.  Here we do not discuss in detail the predictions for many modes,
but focus on a few representative examples.  Discussions about other modes can
be found in many
reviews~\cite{Hitlin:2008gf,Browder:2008em,Bigi:2004kn,Ligeti:2002wt}.

Our first example is extracting $\gamma$ from $B \to D K$. This is
arguably the cleanest measurement in terms of theoretical
uncertainties. Basically, all the necessary hadronic quantities can be
measured. The idea of all $B \to D K$ based
analyses\cite{Gronau:1990ra,Gronau:1991dp,Atwood:1996ci,Giri:2003ty,Grossman:2002aq}
is to consider decays of the type
\beq
B \to D (\ov D)\, K (X) \to f_D\, K (X),
\eeq
where $f_D$ is a final state that is accessible from both $D$ and
$\ov D$ and $X$ represents possible extra particles in the final
state. The crucial point is that, in the intermediate state, the
$D$ or $\ov D$ are not measured (and in particular their flavor
is not tagged), and are thus in a coherent state. On the other hand, 
the $D$ is on-shell, and therefore the
$B\to D$ and the $D\to f_D$ amplitudes are factorized. Thus, we have 
quantum coherence and factorization at the same time.
Using several $B \to DKX$ decays modes (say, $n$
different $X$ states), and several $D \to f_D$ modes (say $k$), one can
perform $n k$ measurements, which depend on $n+k$ hadronic
decay amplitudes. For large enough $n$ and $k$, there is a sufficient
number of measurements to determine all hadronic parameters, as well
as the weak phase we are after. Since all hadronic matrix
elements can be measured, the theoretical uncertainties are much below
the sensitivity of any foreseeable future experiment.

Next in terms of theoretical cleanliness are cases where the leading hadronic
matrix elements cancel. The uncertainties then depend on the ratio of leading and
subleading amplitudes.  The subleading terms can be suppressed by small CKM
matrix elements and/or by loop factors and/or by symmetry breaking corrections
(see examples below).  The question is to estimate the relative size of the
subleading matrix elements.

Consider the time dependent $CP$ asymmetries in neutral $B$ decay
into three different $CP$ eigenstates: $\psi K_S$, $\phi K_S$,
and $\pi^0 K_S$. Using CKM unitarity, Eq.~(\ref{The-UT}), we can write
the relevant decay amplitudes in the form\cite{Grossman:1997gd}
\beq
A = V_{us}\, V_{ub}^*\, A_u\, e^{i \delta} + V_{cs}\, V_{cb}^*\, A_c ,
\eeq
such that $A_i$, which are real and positive, and $\delta$, the $CP$
conserving phase, are mode dependent. The hierarchy of the CKM
elements
\beq
\left|\frac{V_{us}V_{ub}^*}{V_{cs}\, V_{cb}}\right| \sim \lambda^2,
\eeq
suggests that in all three decays we are considering the term 
proportional to $A_u$ is subleading. Indeed, to first approximation, 
this is the case, and all three decays determine $\beta$.
This approximation would fail, however, if $A_u \gg A_c$ due to hadronic 
physics. Even if $A_u\lsim A_c$, we would like to estimate
the corresponding uncertainty, so we need to estimate the ratio
\beq
r \equiv {A_u/A_c}.
\eeq
Here is where hadronic physics enters the analysis, and
the three cases differ.

For $B \to \psi K_S$, the final state contains $c \bar c s \bar d$
quarks, and therefore the quark level decay is dominantly $b \to c
\bar c s$.  Thus, $A_c$ is a tree level decay amplitude, while $A_u$
is a loop (or rescattering) effect.  Consequently, $r \ll 1$ and the
theoretical error in $S_{\psi K_S}$ is
tiny\cite{Gronau:1989ia,Grossman:2002bu,Li:2006vq,Faller:2008zc}. Another
comparable effect proportional to $\epsilon_K$ is due to the fact that
$K_S$ is not a pure $CP$ eigenstate\cite{Grossman:2002bu}.

For $B \to \phi K_S$, the final state contains $s \bar s s \bar d$ quarks, and
therefore the decay is dominantly mediated by $b \to s \bar s s$.  The dominant
contributions to both $A_c$ and $A_u$ come from loop diagrams, so we expect $r
\sim 1$. The theoretical uncertainty in interpreting $S_{\phi K}$ is then
suppressed by $\lambda^2$, and is of order a few
percent\cite{Grossman:1997gr,Grossman:2003qp,Beneke:2005pu}.

For $B \to \pi^0 K_S$, the final state has $u \bar u s \bar d$ quarks and
therefore the quark level decay is $b \to u \bar u s$. Thus, $A_u$ is generated
at tree level while $A_c$ is generated at one loop and we expect $r \gg 1$.
Therefore, the correction to $S_{\pi^0 K_S}$ due to the subleading amplitude
cannot be neglected a-priori. However, it is possible to analyze a set of $B \to
K \pi$ measurements using isospin symmetry relations among them to reduce the
theoretical error on the $CP$ asymmetries.\cite{Gronau:2005kz} The remaining
theoretical uncertainty is at the level of a few percent, due to isospin
breaking. Calculations based on the heavy quark limit also predict small
uncertainty in $S_{\pi^0K_S}$~\cite{Williamson:2006hb,Beneke:2005pu}.

In some cases, approximate symmetries --- isospin or $SU(3)$ --- can be
used to reduce the theoretical uncertainties, allowing clean
extractions of fundamental parameters. The size of $SU(3)$ and $U$-spin 
breaking are comparable, so while we often use only
$U$-spin, we generally refer to it as $SU(3)$. The theoretical errors
associated with these symmetries are at the few percent level for
isospin, and ${\cal O}(20\%)$ for $SU(3)$. Thus, in an era of much
higher precision, when the accuracy of some measurements will be
comparable to isospin breaking, $SU(3)$ may become of limited use.

Isospin has been used in many cases, in particular in $B \to
\pi\pi$\cite{Gronau:1990ka}, $B \to K \pi$, and decays with more pions
in the final states. Isospin is very important in measuring $\alpha$
in decays that proceed via the $b \to u \bar u d$ quark-level
transition, like $B \to \pi\pi$. The theoretical uncertainties arise
from electroweak penguin amplitudes and from isospin breaking in the
matrix elements. The overall theoretical uncertainties are expected to
be at the few percent level\cite{Gardner:1999jq,Falk:2003uq,Gronau:2005pq}.

In $B \to K\pi$, isospin is also crucial in reducing the theoretical
errors. There are many relations between the decay rates and $CP$
asymmetries of the various $B \to K \pi$
modes\cite{Fleischer:1997um,Lipkin:1998ie,Gronau:1998ep}. Here, due to
CKM enhancement, the effect of the electroweak penguin amplitude is
larger than that in $B\to \pi \pi$. Yet, its calculation is considered
reliable and thus precise relations are obtained that can be used to
test the SM.\cite{Neubert:1998pt,Gronau:1998fn} The residual errors
are due to isospin breaking and uncertainties about the magnitude of
the electroweak penguin amplitude. In most cases, we expect the
isospin breaking to enter at first order, and thus to have theoretical
uncertainties at the few percent level. There is, however, one case,
the so-called Lipkin sum rule\cite{Lipkin:1998ie,Gronau:1998ep}, where
isospin breaking affects the result only at second
order\cite{Gronau:2006eb}. Thus, this sum rule has theoretical
uncertainties at the percent level.

Other important theoretical tools come from expanding about the heavy
quark limit, $m_{b(c)} \gg \lqcd$.  There are several formalisms to do
this.  For spectroscopy and exclusive semileptonic decays, extra
symmetries of the Lagrangian emerge in the $m \gg \lqcd$ limit.  These
heavy quark spin-flavor symmetries (HQS)\cite{Isgur:1989vq} imply, for
example, that exclusive semileptonic $B\to D^{(*)}\ell\bar\nu$ decays
are described by a universal Isgur-Wise function in the symmetry
limit, providing some model-independent predictions. For inclusive
semileptonic $B$ decays an operator product expansion (OPE) can be
used to compute sufficiently inclusive
rates~\cite{Chay:1990da,Bigi:1992su,Manohar:1993qn}.  The leading
order result is given by free quark decay, the $\lqcd/m_b$ terms
vanish, and the $\lqcd^2/m_b^2$ corrections are parameterized by just
two hadronic matrix elements, which can be determined from data. Thus,
the theoretical uncertainties for inclusive semileptonic rates are at
the few percent level.  A prime application is the extraction of
$|V_{cb}|$; the theoretical uncertainties are at the few percent level
both in the inclusive and exclusive analysis.  When severe phase space
cuts are imposed experimentally, such as for many determinations of
$|V_{ub}|$ from inclusive decays, the expansion is less powerful,
since the usual OPE in terms of local matrix elements is replaced by a
``nonlocal OPE" in which the hadronic matrix elements are functions
rather than numbers, and $\lqcd/m_b$ corrections do occur.  The heavy
quark expansion in fully hadronic decays~\cite{Beneke:1999br} becomes
yet more complicated, and motivated many developments in
soft-collinear effective theory (SCET)~\cite{Bauer:2000ew}.  In most
such applications, we do not have a complete categorization of even
the leading power corrections. Furthermore, the corresponding matrix
elements have substantial uncertainties. Thus, the theoretical
uncertainties are at least at the $10\%$ level, and often larger. One
of the possible exceptions is the difference of $CP$ asymmetries,
$A_{K^+\pi^0} - A_{K^+\pi^-} = 0.15 \pm 0.03$, which appears hard to
reconcile with the heavy quark expansion and any set of assumptions
about the $\lqcd/m_b$ corrections popular in the literature.

A theoretical tool where significant improvements are expected in the
next few years is lattice QCD.  In principle, lattice QCD enables us
to calculate many nonperturbative matrix elements.~\cite{Gamiz:2008iv}
In practice, however, several approximations have to be used to keep
the computational time under control, e.g., because the $b$ quark is
too heavy to be simulated directly.  Yet, we can hope to see
improvements in the next few years as new algorithms and more powerful
computers are used.  One may hope that matrix elements which contain
at most one (stable) hadron in the final state may be calculated with
percent level uncertainties.  Matrix elements involving states with
sizable widths, e.g., $\rho$ and $K^*$, are more challenging.  Matrix
elements containing more than one hadron in the final state are much
more complicated, and it would require major developments to be able
to do calculations with small and reliable uncertainties.  Thus, the
theoretical errors are expected to shrink especially for measurements
that relate to meson mixing, leptonic and semileptonic decays.


While our main focus is $B$ physics, there are interesting and
theoretically clean observables in $K$ and $D$ decays. In particular,
the decay rates of $K^+ \to \pi^+ \nu \bar\nu$ and $K_L \to \pi^0 \nu
\bar\nu$ are very clean, with theoretical errors at the few percent
level\cite{Buchalla:1994tr,Buchalla:1996fp,Grossman:1997sk,Buras:2005gr}. 
In the neutral $D$ system, the calculations of the mass and width
differences suffer from large hadronic
uncertainties.\cite{Georgi:1992as,Bigi:2000wn,Falk:2001hx,Falk:2004wg}
Yet, the general prediction of the SM is that all $CP$ asymmetries in
tree level decays are below the $10^{-2}$ level. The reason is that
charm decay and mixing involve to a good approximation only the two lighter
generations, and are thus insensitive to the SM $CP$
violation. Consequently, if $CP$ violation is observed in charm mixing
or decay, it will be a good probe of new flavor
physics.\cite{Blaylock:1995ay,Grossman:2006jg,Golowich:2007ka}

Finally, we mention lepton flavor violating decays, such as $\tau \to
\mu\mu\mu$.  Such decays can be studied in future $B$ factories, and are very
clean theoretically. The SM prediction is zero for all such decays. Adding
neutrino masses to the SM via the see-saw mechanism yields lepton flavor
changing operators suppressed by the see-saw scale.  The resulting lepton flavor
violating branching ratios are tiny, many orders of magnitude below the
experimental sensitivities. Thus, these decays are very clean probes of
TeV-scale physics.

Our conclusion is that there are many observables with theoretical
uncertainties at the few percent level, and some with even smaller
errors. This is an important conclusion, since many of these observables
are expected to be measured with a percent-level accuracy, which will 
allow us to discover small contributions from new physics. As argued 
above, this is also the level of deviations
expected from many interesting new physics models.

\section{Expected sensitivity to deviations from the SM}
Most of the currently available $B$ decay data is from the two $B$
factories, \babar\ and Belle, and from the Tevatron experiments, CDF
and D\O. Not much more data is expected from the $B$ factories:
\babar\ finished its data taking and Belle will stop running in its
current configuration soon. While CDF and D\O\ are still running, the
expected increase in integrated luminosity is at most a factor of
few. Much more data is expected to be accumulated by future
experiments. First, LHCb will start to operate soon.  They expect
10\,fb$^{-1}$ of data collected by 2015 or so. Beyond that, an LHCb
upgrade is planned with 10 times larger luminosity. This hadron
collider data is complementary to the $e^+e^-$ data. The statistics is
much higher and the sensitivity to the various decay modes is
different. There are also proposals for higher luminosity $e^+e^-$
machines. One proposal is to upgrade KEK-B to reach a luminosity near
$10^{36}/{\rm cm}^2/{\rm s}$. The other proposal is to build a new
machine in Italy with a luminosity of $10^{36}/{\rm cm}^2/{\rm s}$ or
possibly even higher.

Any attempt to assess the sensitivity of future measurements, 
with a factor of 100 larger statistics than currently available, 
is unavoidably subject to significant uncertainties.  For example, 
for the expected super-$B$-factory sensitivities,
various studies~\cite{Akeroyd:2004mj,Hewett:2004tv,Bona:2007qt} assume quite
different beam conditions and detectors.  In Table~\ref{tab:future} we list the
current status~\cite{Barberio:2008fa} and expected sensitivities in some of the
channels we view as important.  The LHCb expectations are taken from
Refs.~\cite{Buchalla:2008jp,Altarelli:2008xy,LHCb}. We emphasize that this
table is not comprehensive and the entries in the last two columns necessarily 
have significant uncertainties. 

\begin{table}[t]
\begin{tabular}{ccc|cc}\hline
\raisebox{-6pt}{Observable}  &  Approximate  &  Present  &
  \multicolumn{2}{c}{Uncertainty / number of events} \\[-2pt]
  &  SM prediction  &  status
  &  Super-$B$ (50\,ab$^{-1}$)  &  LHCb (10\,fb$^{-1}$) \\ \hline\hline
$S_{\psi K}$  &  input  &  $0.671 \pm 0.024$
  &  0.005  &  0.01 \\
$S_{\phi K}$  &  $S_{\psi K}$  &  $0.44 \pm 0.18$
  &  0.03  &  0.1 \\
$S_{\eta' K}$  &  $S_{\psi K}$  &  $0.59 \pm 0.07$
  &  0.02  &  not studied \\
$\alpha(\pi\pi,\rho\rho,\rho\pi)$  &  $\alpha$  &  $(89\pm4)^\circ$
  &  $2^\circ$  &  $4^\circ$ \\
$\gamma(DK)$  &  $\gamma$  &  $(70^{+27}_{-30})^\circ$
  &  $2^\circ$  &  $3^\circ$ \\
$S_{K^* \gamma}$  &  few $\times$ 0.01  &  $ -0.16 \pm 0.22$
  &  $0.03$  &  --- \\
$S_{B_s\to\phi\gamma}$  &  few $\times$ 0.01  &  ---
  &  ---  &  $0.05$ \\
$\beta_s(B_s\to \psi\phi)$  &  $1^\circ$  &
  $(22^{+10}_{-8})^\circ$  &  ---  &  $0.3^\circ$ \\
$\beta_s(B_s\to \phi\phi)$  &  $1^\circ$  &  ---
  &  ---  &  $1.5^\circ$ \\
$A_{\rm SL}^d$  &  $-5 \times 10^{-4}$  &  $-(5.8 \pm 3.4) \times 10^{-3}$
  &  $10^{-3}$  &  $10^{-3}$ \\
$A_{\rm SL}^s$  &  $2 \times 10^{-5}$  &  $(1.6 \pm 8.5) \times 10^{-3}$
  &  $\Upsilon(5S)$ run?  &  $10^{-3}$ \\
$A_{CP}(b\to s\gamma)$  &  $<0.01$  &  $-0.012\pm0.028$  &  0.005  & ---\\[2pt]
\hline \multicolumn{3}{c|}{}\\[-9pt]
$|V_{cb}|$  &  input  &  $(41.2 \pm 1.1)\times 10^{-3}$
  &  1\%  & --- \\
$|V_{ub}|$  &  input  &  $(3.93 \pm 0.36)\times 10^{-3}$
  &  4\%  & --- \\
$B\to X_s\gamma$  &  $3.2\times 10^{-4}$  &  $(3.52 \pm 0.25)\times 10^{-4}$
  &  4\%  & --- \\
$B\to \tau\nu$  &  $1\times 10^{-4}$  &  $(1.73\pm0.35)\times10^{-4}$
  &  5\%  & --- \\
$B\to X_s\nu\bar\nu$  &  $3\times 10^{-5}$  &  $<6.4\times10^{-4}$
  &  only $K\nu\bar\nu$\,?  & --- \\
$B\to X_s \ell^+ \ell^-$  &  $6\times 10^{-6}$  &  $(4.5\pm1.0)\times10^{-6}$
  &  6\%  &  not studied\\
$B_s\to \tau^+\tau^-$  &  $1\times 10^{-6}$  &  $<$ few \%
  &  $\Upsilon(5S)$ run?  &  --- \\
$B\to X_s\,\tau^+\tau^-$  &  $5\times 10^{-7}$  &  $<$ few \%
  &  not studied  &  --- \\
$B\to \mu\nu$  &  $4\times 10^{-7}$  &  $<1.3\times10^{-6}$
  &  6\%  & --- \\
$B\to \tau^+\tau^-$  &  $5\times 10^{-8}$  &  $<4.1\times 10^{-3}$
  &  ${\cal O}(10^{-4})$  &  --- \\
$B_s\to \mu^+\mu^-$  &  $3\times 10^{-9}$  &  $<5\times10^{-8}$
  &  ---  &  $>5\sigma$ in SM \\
$B\to \mu^+\mu^-$  &  $1\times 10^{-10}$  &  $<1.5\times10^{-8}$
  &  $< 7 \times 10^{-9}$  &  not studied \\[2pt]
\hline \multicolumn{3}{c|}{}\\[-9pt]
$B\to K^*\ell^+\ell^-$  &  $1\times 10^{-6}$  &  $(1\pm 0.1)\times 10^{-6}$
  &  15k  &  36k\\
$B\to K\nu\bar\nu$  &  $4\times 10^{-6}$  &  $< 1.4\times 10^{-5}$
  &  20\%  &  ---
\\ \hline
\end{tabular}\vspace*{4pt}
\caption{Some interesting observables.  In the ``present status" column, upper
bounds are 90\%\,CL.  The expected experimental sensitivities are current
estimates and may change in the future.  In several processes the most
interesting information will come from more detailed measurements that cannot be
captured simply by a single number.}
\label{tab:future}
\end{table}

The simplest extrapolations can be done for those measurements in
which the experimental uncertainty is expected to be dominated by
statistical errors for any foreseen data sets, and the theoretical
uncertainties are negligible.  An example of this is the 
determination of the CKM angle $\gamma$.  In other cases, 
theoretical uncertainties may still be very small, but the experimental 
systematic errors become important. An example of this is $A_{\rm SL}^{d,s}$.  
In less favorable modes, systematic errors may become dominant on both the
experimental and the theoretical sides.  Interestingly, the
gold-plated measurement of \babar\ and Belle, $S_{\psi K}$, is in this
category. None of the
reports~\cite{Akeroyd:2004mj,Hewett:2004tv,Bona:2007qt} expect the
uncertainty to decrease by a factor of 10 with 100 times more data.
As discussed above, at the 0.005 error level, the relation
between $S_{\psi K}$ and $\sin2\beta$ is sensitive to hadronic
physics. Similar is the determination of $\alpha$, where some of the
uncertainties not yet addressed in the current analysis (e.g., isospin 
violation) may become relevant when the experimental precision improves.
To what extent they can be controlled using the data~\cite{Falk:2003uq} 
is hard to foresee.

The magnitudes of CKM elements are important for constraining new physics by
comparing the information from tree-dominated and loop-mediated processes.  Some
$|V_{cb}|$ and $|V_{ub}|$ analyses are already theory limited, while some of the
theoretically cleaner (and experimentally less efficient) methods will benefit
from more data. In particular, the experimental implementation of many of the
theoretically cleaner analyses is based on the full-reconstruction tag method,
in which the ``other" $B$ meson is fully reconstructed.  This way, the
four-momenta of both the leptonic and the hadronic systems can be measured. It
also gives access to a wider kinematic region due to improved signal purity, and
is only possible in the $e^+e^-$ environment.  The possibility to precisely
determine $|V_{cb}|$ and $|V_{ub}|$ from exclusive decays is almost entirely in
the hands of lattice QCD.  While a lot of progress is expected, we will need
experience to assess under what circumstances one can prove the presence of new
physics if there is significant tension between data and lattice QCD predictions
(e.g., how  the $f_{D_s}$-problem~\cite{Follana:2007uv} is going to be
resolved).  At present, there is also some tension between the inclusive and
exclusive measurements of both $|V_{cb}|$ and $|V_{ub}|$, which prompted the PDG
in 2008 for the first time  to inflate the errors~\cite{Amsler:2008zzb}.

Many rare FCNC decays are sensitive probes of various extensions of the SM.  In
most cases the theory is under better control for inclusive decays, which are
very hard (if not impossible) to measure at hadron colliders.  Final states with
neutrinos or $\tau$ leptons producing final states with large missing
energy~\cite{Grossman:1995gt,Grossman:1996qj,Barate:2000rc} are only possible at
$e^+e^-$ colliders.  At the same time, the constraints on the very important
$B_s\to\mu^+\mu^-$ and $B\to\mu^+\mu^-$ modes are, and will be, dominated by
hadron collider data.

The two exclusive decay modes at the bottom of the table are listed
for the following reasons. The $K\nu\bar\nu$
mode~\cite{Altmannshofer:2009ma} may be the only final state in $B\to
X_s\nu\bar\nu$ decays that can be measured with good precision.  For
the $K^*\ell^+\ell^-$ mode, there are many interesting differential
observables, such as the zero of the forward-backward asymmetry, which
LHCb expect to measure with a $0.3\,\GeV^2$ uncertainty with
10\,fb$^{-1}$ data, giving a determination of $C_7^{\rm eff}/C_9^{\rm
eff}$ with 7\% statistical error.

The role of LHCb is even more important than Table~\ref{tab:future} might
indicate.  First, at this time, it is the only future dedicated $B$ physics
experiment which will definitely take data.  Its expected measurement of
$B_s\to\mu^+\mu^-$ is particularly sensitive to some extensions of the SM not
well constrained so far.  Furthermore, if we parameterize new physics in
$B_s^0$--$\B0bar_s$ mixing with $h_s$ and $\sigma_s$, similar to
Eq.~(\ref{hsigma}), then the measurement of $\beta_s$ will give constraints
similarly strong as those on $h_d,\sigma_d$.~\cite{Ligeti:2006pm}

While we concentrate on the future of $B$ physics, a super-$B$-factory will also
be sensitive to other kinds of physics. It will probe lepton flavor violating
decays at a much improved level.  For example, sensitivity down to ${\cal
B}(\tau \to \mu\gamma) \sim 2\times 10^{-9}$ may be achieved. The corresponding
bound on the ratio ${\cal B}(\mu \to e\gamma) / {\cal B}(\tau \to \mu\gamma)$
will be useful to constrain various new physics models. Similarly, decays of the
type $\tau^- \to \ell_1^-\ell_2^-\ell_3^+$ will also be constrained at a level
around ${\cal B}(\tau \to \mu\mu\mu) \sim 2 \times\, 10^{-10}$.  Again, the
ratio ${\cal B}(\tau \to \mu\mu\mu) / {\cal B}(\tau \to \mu\gamma)$ is an
interesting probe of new physics; for example, if new physics generates the
operators $\bar\tau_R\sigma_{\alpha\beta} F^{\alpha\beta} \mu_L$ and
$(\bar\tau_L \gamma^\alpha\mu_L) (\bar\mu_L \gamma_\alpha \mu_L)$ with
coefficients of very different magnitudes, then either decay mode can be more
sensitive to a particular model. A super-$B$-factory will also be able to do
precision QCD studies and look for yet unobserved resonances. It may even be
able to find direct signals of new physics, such as a very light Higgs boson or
light particles\cite{Aubert:2009cp} predicted in some models of dark
matter.\cite{ArkaniHamed:2008qn,Nomura:2008ru,Batell:2009yf,Essig:2009nc}

\section{Synergy and complementarity with LHC new particle searches}

The LHC will soon start its operation. As can be learned from
Table~\ref{tab:future}, the LHCb experiment will contribute to our
understanding of flavor via measurements of $B$ and $B_s$ decays.
ATLAS and CMS will also probe FCNC top quark decays at orders of magnitude
better level than the current bounds~\cite{Carvalho:2007yi,Fox:2007in}.
There is, however, another aspect where the LHC is expected to be
relevant to flavor physics, and that is the interplay between the
high-$p_T$ physics of the ATLAS and CMS detectors and the low energy
measurements of the flavor factories.

Let us first comment on a rather pessimistic scenario, where no new
physics is observed at ATLAS/CMS. In that case, we will lose the two
main clues that we have had to new physics at the TeV scale. First, we
have probably misinterpreted the fine-tuning problem of the Higgs
mass. Second, the dark matter particles are perhaps not WIMPs. It will
be difficult then to argue that a collider with a center of mass
energy of, say, 50 or 100 TeV, is likely to discover new physics. It
is this point where flavor physics might play an important role.
If deviations from the standard model predictions for FCNC processes
are established in the flavor factories, they can be used to put an
upper bound on the scale of new physics. This might be then the only
argument for a relatively low energy new physics. Depending on the
pattern of deviations, one might get further clues about the nature of
this new physics, and perhaps propose an experimental program that
can directly produce and study the new physics.

A much more exciting scenario is one where new physics is observed at
ATLAS and CMS. In this case, the interplay between collider physics
and flavor physics is expected to be very fruitful:
\begin{itemize}
\item It is very likely that we will understand how the new physics
flavor puzzle is solved, namely what are the special flavor features
of the new physics that are at work in suppressing its contribution to
FCNC processes.
\item Under some favorable circumstances, we may also get clues about
  the solution to the standard model flavor puzzle, namely why there
  is hierarchy and smallness in the Yukawa couplings.
\item Understanding the flavor structure of the new physics might
  teach us about its inner structure and perhaps about physics at a
  scale much higher than the LHC scale.
\end{itemize}
In the rest of this section, we explain these three points in more
detail.

If new particles are discovered at the LHC, and if they couple to the
standard model quarks and/or leptons, then there are new flavor
parameters that can, at least in principle, be measured. These include
the spectrum of the new particles, and their flavor decomposition, {\it
i.e.}, their decay branching ratios. Realistically one can expect
that ATLAS/CMS will be able to measure the leptonic flavor
decomposition and, for the quark sector, separate third generation
final states (bottom and top) from first two generation final states.
Even this limited information might complement in a significant way
the information on flavor from FCNC measurements.

As an example of how this information will allow us to make progress
in solving the new physics flavor puzzle, consider the principle of
minimal flavor violation (MFV)
\cite{D'Ambrosio:2002ex,Hall:1990ac,Chivukula:1987py,Buras:2000dm}.
Within the standard model, there is a large global symmetry,
$[U(3)]^5$, which is broken only by the Yukawa matrices. Models of new
physics where this is still true, namely where there are no new
sources of flavor violation beyond the standard model Yukawa matrices,
are said to obey the principle of MFV. A well known example of an MFV
model is that of gauge mediated supersymmetry breaking. The scale of
MFV models can be as low as order TeV without violating the bounds
from FCNC, thus solving the new physics flavor puzzle.

Within models of MFV, the only quark flavor changing parameters are
the CKM elements. Since the CKM elements that couple the third
generation to the lighter ones are very small, new particles in MFV
models that decay to a single final quark decay to either a third
generation quark or to a first two generation quark, but (to ${\cal
  O}(10^{-3})$ approximation) not to both \cite{Grossman:2007bd}. As a
concrete example, consider extra heavy quarks in vector-like
representations of the standard model gauge group. If ATLAS/CMS
discover such a heavy quark and can establish that it decays to both
third generation and light quarks, it will prove that MFV does not
hold and is not the solution to the new physics flavor puzzle
\cite{Grossman:2007bd}.

Within supersymmetry, the predictive power of MFV is even stronger. It
implies that the first two squark generations are quasi-degenerate and decay
only to light quarks, while the third generation squarks are separated
in mass (except for $SU(2)$-singlet down squarks if $\tan\beta$ is
small) and decay to only third generation quarks \cite{Nir:2007xn}. If
these features can be tested, it will provide us not only with deeper
understanding of how the supersymmetric flavor puzzle is solved, but
also open a window to the mechanism that mediates supersymmetry
breaking, which is physics at a scale much higher than those
accessible to the LHC \cite{Feng:2007ke}.

The standard model flavor parameters -- fermion masses and mixing
angles -- seem to have a structure. In particular, there is smallness
and hierarchy in the Yukawa sector. An explanation to this structure
might lie in some high scale physics, perhaps an approximate symmetry,
such as the Froggatt-Nielsen mechanism \cite{Froggatt:1978nt}, or a
dynamical mechanism, such as the Nelson-Strassler mechanism
\cite{Nelson:2000sn}. If new physics is discovered at the LHC, then we
might ask whether its flavor structure is determined by the same
mechanism as the Yukawa couplings. For example, the mixing among
sleptons in a certain class of supersymmetric models can test the
Froggatt-Nielsen mechanism\cite{Feng:2007ke}.

\begin{figure}[t!]
\centerline{\includegraphics*[height=7.5cm]{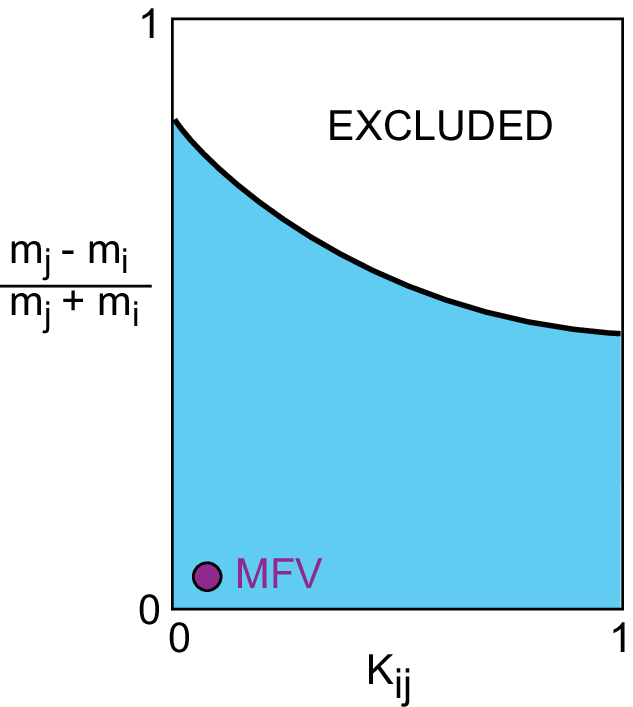} \hfil
\includegraphics*[height=7.5cm]{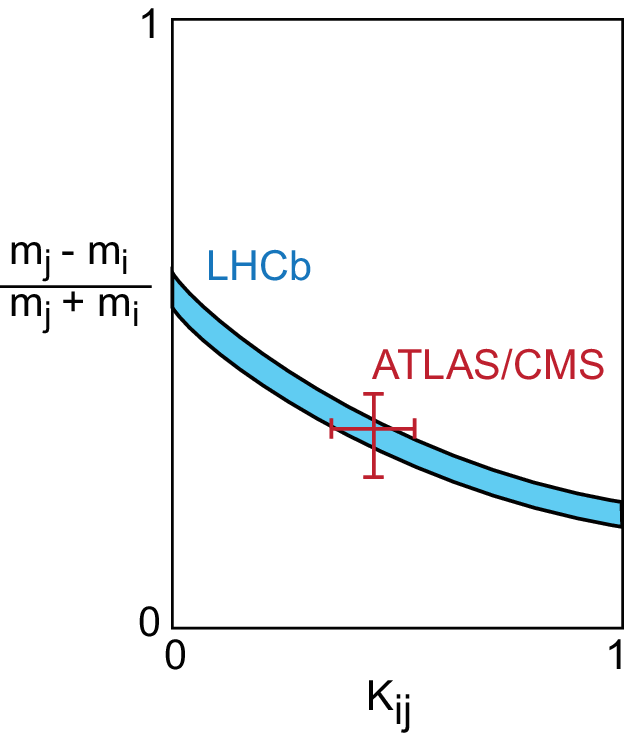}}
\vspace*{-.25cm}
{\footnotesize\hspace*{3cm} (a)\hspace*{65mm}(b)}
\caption{A schematic description of the constraints on the mass
  splitting, $(m_i-m_j)/(m_i+m_j)$, and mixing angle, $K_{ij}$,
  between squarks (or sleptons). Left: A typical present constraint
  arising from not observing deviations
  from the standard model predictions. The fact that the region of
  splitting and mixing of order one is excluded constitutes the new
  physics flavor puzzle. MFV often implies deviations that are much
  smaller than present sensitivities. Right: A possible future scenario
  where the mass splitting and the flavor decomposition are measured
  by ATLAS/CMS, and they fit deviations observed in a flavor factory.}
\label{fig:dmk}
\end{figure}

The only information that we have at present on flavor aspects of
possible new physics models is coming from low energy
measurements. Such measurements put constraints on the product of the
flavor alignment and flavor degeneracy factors (see, however,
\cite{Blum:2009sk}). For example, within
supersymmetry, measurements of $b\to s$ transitions constrain the
product of the mass splitting between the second and third generation
down squarks, and their mixing. If in the future we discover the new
physics, then we can in principle measure the flavor alignment and
flavor degeneracy separately. If, in addition, future $B$ factories
establish a deviation from the standard model for such a process, then
the overall consistency of the direct measurements and the low energy
measurement will assure us that we fully understand the flavor
structure of the new physics and how the new physics flavor puzzle is
solved. The present situation and an (optimistic) future scenario are
schematically depicted in Fig.~\ref{fig:dmk}.

\section{Conclusions}

We are now in a transition period in flavor physics. In the past,
flavor physics led the way to the three generation SM, with the
CKM picture of flavor and $CP$ violation. The 2008 Nobel Prize 
in Physics, awarded to Kobayashi and Maskawa, is a formal recognition 
that this task is, to large extent, completed. We know that the CKM
matrix is the dominant source of flavor and $CP$ violation.

Here we discussed the future of flavor physics, in particular, that of
$B$ physics. The question at hand is not less important: Given that the
SM is only a low energy effective theory that is very likely to be
supplemented by new physics at a scale close to TeV, how can flavor
physics help in understanding the ultraviolet completion of the SM? 
The fine tuning problem of the Higgs mass and the dark matter
puzzle lead us to think that there is new physics at the TeV scale.
Flavor bounds tell us, in turn, that such new physics must have
a special flavor structure.

To make sure that the future of flavor physics may be as successful
as its past, we need much more data. Present constraints imply
that the new physics, while perhaps not minimally flavor violating, is
likely to have flavor suppression factors that are similar to, or
stronger than, the SM ones. Since the new physics contributions are 
further suppressed by a scale
somewhat above the electroweak scale, we expect to see small deviations
from the SM prediction. A more quantitative statement would be both
model dependent and mode dependent, but the main idea is clear: ``small
deviations" mean that more data is required to discover
them. Conversely, with more data, it is not unlikely that
deviations will indeed be discovered.

Three facts combine to make the future look promising:
\begin{itemize}
\item The technology to collect much more flavor data exists. This includes 
the LHC experiments that will operate in a hadron environment, and
the proposed super-$B$-factories that will be high luminosity
$e^+e^-$ machines.
\item Many measurements are not theory limited. That is, the new data can be
compared to solid theoretical predictions. 
\item In many modes we expect deviations to be found at the level of the
experimental sensitivity and above the SM theoretical errors.
\end{itemize}
We may thus be optimistic about 
obtaining convincing  pieces of evidence for new physics in the flavor sector. 
Together with the anticipated direct discovery of new particles at the LHC, 
we will be able to learn a great deal about the way that Nature works 
at a very fundamental level.

\subsection*{\bf Acknowledgements}
We thank Tatsuya Nakada and Jesse Thaler for helpful discussions.  
This work is supported by the United States-Israel Binational Science
Foundation (BSF), Jerusalem, Israel.  The work of YG is supported by
the NSF grant PHY-0757868. The work of ZL was supported in part by the
Director, Office of Science, Office of High Energy Physics of the
U.S.\ Department of Energy under contract DE-AC02-05CH11231.  The work
of YN is supported by the Israel Science Foundation founded by the
Israel Academy of Sciences and Humanities, the German-Israeli
foundation for scientific research and development (GIF), and the
Minerva Foundation.

\end{document}